\documentclass[aps,prb,reprint,superscriptaddress,floatfix,longbibliography]{revtex4-2}
\usepackage{amsmath,amssymb,bm}
\usepackage{graphicx}
\usepackage{xcolor}
\usepackage{booktabs}
\usepackage{siunitx}
\usepackage{microtype}
\usepackage[hidelinks]{hyperref} 
\graphicspath{{figures/}}
\begin{document}

\title{Probing Chern Topology through Hall-Sensitive Photonic Spin Hall Shifts in a $d$-wave Altermagnet}

\author{Mahnoor Sajid}
\affiliation{Department of Physics, Quaid-i-Azam University, Islamabad,  Pakistan}
\author{Maha Siddiqui}
\affiliation{Department of Physics, Quaid-i-Azam University, Islamabad,  Pakistan}
\author{Muzamil Shah}
\email{muzamil@qau.edu.pk}
\affiliation{Department of Physics, Quaid-i-Azam University, Islamabad,  Pakistan}
\date{\today}

\begin{abstract}
In this paper, we investigate the photonic spin Hall effect in a two-dimensional Rashba-coupled $d$-wave altermagnetic Chern system subjected to a uniform out-of-plane exchange field. We first analyze the
energy-band structure and the Dirac mass distribution
at the high-symmetry points, where
$m_\Gamma=m_M=\Delta_z$,
$m_X=\Delta_z-4t_J$, and $m_Y=\Delta_z+4t_J$. Band inversions at high-symmetry points generate both a topologically trivial regime ($C=0$) and Chern-insulating phases ($C=\pm 1$). Employing the Kubo formalism for the optical conductivity together with the reflection matrix for a substrate-supported conducting sheet, we compute helicity-dependent finite-waist displacements of a reflected Gaussian beam. The resulting optical response exhibits a minimum interband absorption threshold, a Van Hove–type spectral feature, and distinct transitions at the $X$ and $Y$ points arising from the momentum dependence of the exchange field. Under the specified Hall-tensor and polarization sign conventions, the longitudinal helicity half-separation is odd upon reversal of the optical Hall conductivity and correspondingly changes sign between opposite-Chern phases, whereas the transverse half-separation is even in the Hall response. The maximal spin-dependent splitting occur in the vicinity of the pseudo-Brewster reflectance minimum. The longitudinal spin splitting reverses sign
between opposite Chern phases, reflecting the
exchange-controlled optical Hall response,
while its pronounced spectral variations reveal
the influence of inequivalent interband
transitions induced by the $d$-wave altermagnetic
exchange. These findings establish
polarization-resolved photonic beam shifts as
a sensitive probe of Hall-active
topological phases and their characteristic
optical spectra.
\end{abstract}

\maketitle

\section{Introduction}
\label{sec:intro}

Altermagnetism has emerged as a distinct form of
collinear magnetic order that combines the compensated
magnetic structure of conventional antiferromagnets
with momentum-dependent spin splitting reminiscent
of ferromagnetic systems~\cite{Smejkal2022PRX,Smejkal2022Landscape,Mazin2022}.
In altermagnets, oppositely polarized magnetic
sublattices are connected by crystal rotations
rather than by translations or inversion operations,
giving rise to unconventional $d$-, $g$-, or
$i$-wave spin-splitting patterns in momentum
space~\cite{Hayami2019,Smejkal2022PRX}.
Unlike conventional spin--orbit-driven band
splitting, the characteristic altermagnetic
spin splitting can originate from nonrelativistic
exchange interactions and persist even in
the absence of spin--orbit coupling~\cite{Bai2024,Song2025}.
The compensated nature of the magnetic order
suppresses macroscopic stray fields while
preserving nontrivial spin-dependent electronic
properties, making altermagnets attractive for
spintronics and the investigation of unconventional
transport phenomena.
Experimental evidence for altermagnetic band
splitting has been obtained through photoemission
measurements in MnTe~\cite{Krempasky2024,Lee2024,Osumi2024} and
CrSb~\cite{Reimers2024,Ding2024}, while a room-temperature
metallic $d$-wave altermagnetic state has been
reported in layered KV$_2$Se$_2$O~\cite{Jiang2025}.

One of the particularly intriguing properties
of altermagnetic systems is their ability to
support unconventional transverse charge
transport without requiring a conventional
ferromagnetic state.
Although the anomalous Hall effect (AHE) \cite{RevModPhys.82.1539}
is commonly associated with ferromagnetic
materials, where broken time-reversal symmetry
and spin--orbit coupling generate a finite
Berry-curvature contribution to the transverse
conductivity, an analogous response can also
emerge in certain compensated magnetic
systems when allowed by their magnetic
crystal symmetry~\cite{Smejkal2022Landscape,GonzalezBetancourt2023,Reichlova2024}.
In particular, anomalous Hall responses
have been experimentally investigated in
MnTe and Mn$_5$Si$_3$, including measurements
in the absence of an externally applied
magnetic field~\cite{GonzalezBetancourt2023,Reichlova2024}.
These observations demonstrate that a
macroscopic ferromagnetic magnetization
is not always required for an anomalous
Hall response.
However, the existence and magnitude of
the Hall conductivity in an altermagnet
depend critically on the underlying
magnetic symmetries, spin--orbit coupling,
and the orientation of the N\'eel vector.
Consequently, the anomalous Hall response
may be allowed or forbidden depending on
the particular magnetic configuration.
This strong connection between magnetic
symmetry and Berry-curvature-driven
transport motivates the search for
optical methods capable of identifying
Hall-active electronic states and
distinguishing their topological properties.

The optical counterpart of the anomalous
Hall effect provides an important route
toward such characterization.
The complex optical conductivity tensor,
with longitudinal component
$\sigma_{xx}(\omega)$ and transverse
component $\sigma_{xy}(\omega)$,
encodes the interband transition spectrum
and the Berry-curvature-dependent
electromagnetic response.
The longitudinal conductivity reflects
the available optical transitions,
whereas the transverse conductivity
contains information about broken
time-reversal symmetry and the associated
Hall response.
In insulating topological phases,
the low-frequency Hall conductivity
is directly related to the Chern number,
while finite-frequency structures
reflect the underlying electronic
dispersion and optical selection rules.
These responses can be investigated
through conventional magneto-optical
observables, including Faraday and
Kerr rotations~\cite{Tse2010,Tse2011}.
Recent theoretical investigations
have also explored the optical and
magneto-optical properties of
altermagnetic systems, highlighting
the influence of magnetic symmetry,
spin--orbit coupling, and band
topology on their electromagnetic
response~\cite{Rao2024,Chen2025,Iguchi2025,LiLiu2025,Tagani2026}.
An alternative possibility is to
probe these conductivity-dependent
features through the photonic spin-dependent shifts.

The photonic spin Hall effect (PSHE)
offers a sensitive optical approach
for investigating the interaction
between light and condensed-matter
systems.
It originates from the spin--orbit
interaction of light, through which
the polarization degree of freedom
becomes coupled to the spatial
trajectory of an optical beam
~\cite{Onoda2004,Bliokh2006,Hosten2008,Bliokh2015}.
Upon reflection or refraction at
an optical interface, the two
opposite circularly polarized
components of an incident beam
can experience different spatial
or angular displacements.
The transverse spin-dependent
displacement is closely associated
with the Imbert--Fedorov effect,
whereas longitudinal shifts
are related to the
Goos--H\"anchen effect
~\cite{BliokhAiello2013,Ling2017}.
These shifts depend on the
amplitudes, phases, and angular
derivatives of the Fresnel
reflection coefficients and
can be significantly enhanced
near the Brewster angle,
where the ordinary reflected
amplitude becomes small
~\cite{Luo2011,Kong2012}.
Because the Fresnel coefficients
are sensitive to the optical
properties of the reflecting
medium, the PSHE provides
a potentially useful probe of
electronic and optical responses.

Considerable attention has been
devoted to the PSHE in a variety
of material systems, including
conventional dielectric interfaces,
graphene, two-dimensional
semiconductors, and topological
materials~\cite{BliokhAiello2013,Luo2011,Kong2012,Zhou2012,KortKamp2016,Cai2017,KortKamp2017,Shah2022}.
In graphene-based structures,
the dependence of the reflected
beam displacement on the sheet
conductivity has been explored
as a means of probing the
properties and number of
atomic layers~\cite{Zhou2012}.
Related investigations in quantum
Hall systems have established
connections between Hall
conductivity and spin-dependent
optical shifts~\cite{KortKamp2016,Cai2017}.
Moreover, PSHE studies in
two-dimensional topological
materials have demonstrated
that electronic band topology
can strongly influence
the polarization-dependent
reflection response.
In particular, Kort-Kamp
investigated topological
phase transitions through
photonic spin Hall shifts
in graphene-family materials
~\cite{KortKamp2017,Shah2022}.
Shah \textit{et al.}~\cite{Shah2024}
subsequently studied the
longitudinal and transverse
spin-dependent displacements
in Haldane-model systems,
buckled two-dimensional
materials, and transition
metal dichalcogenides,
demonstrating their sensitivity
to topological phases as
well as spin- and valley-dependent
electronic properties.
More recently, Samanta and
Joshi~\cite{Samanta2026} examined
the relationship between
the optical Hall response
and the intensity-weighted
transverse beam displacement.
Together, these studies
establish the PSHE as a
promising optical probe of
two-dimensional electronic
systems and motivate its
extension to magnetic
materials with unconventional
momentum-dependent spin
structures.

Despite these developments,
the relationship between
altermagnetic band structures,
topological Hall conductivity,
and helicity-resolved optical
beam shifts remains less
explored.
Altermagnetic systems are
particularly interesting
because their momentum-dependent
exchange fields can modify
the electronic dispersion,
Berry curvature, and
interband optical transitions
without relying on the
conventional ferromagnetic
spin-splitting mechanism.
When combined with spin--orbit
coupling and suitable
symmetry-breaking perturbations,
these features can support
topologically nontrivial
electronic phases
~\cite{Ma2024,LiLiu2024,LiZhang2025,Antonenko2025,Parshukov2025,GonzalezHernandez2025,Rao2024,Cheraghchi2026}.
The resulting optical
conductivity can therefore
contain distinct signatures
of magnetic exchange,
band inversion, and
topological phase formation.
Since the conductivity tensor
directly influences the
polarization-dependent
reflection coefficients,
these signatures may also
appear in the photonic
beam shifts.
It is consequently important
to determine which displacement
observables are sensitive
to the sign of the Hall
response, how they differ
between trivial and
Chern-insulating phases,
and whether the interband
spectral features can be
resolved through reflected
beam displacements.

\begingroup\color{black}
To address these questions, we consider a two-dimensional
$d$-wave altermagnetic Chern model incorporating Rashba
spin--orbit coupling, momentum-dependent altermagnetic
exchange, and a uniform out-of-plane exchange field.
The interplay of these interactions produces topologically
trivial and nontrivial insulating phases with Chern numbers
$C=0,\pm1$.  In this work, we investigate the longitudinal and transverse
photonic spin-dependent displacements in the altermagnetic
system by incorporating its Kubo optical conductivity into
the generalized Fresnel reflection coefficients. We show
that the longitudinal spin splitting is sensitive to reversal
of the optical Hall conductivity, providing a means of
distinguishing opposite Chern phases, whereas the transverse
splitting exhibits a Hall-even response. Furthermore,
pronounced spectral features emerge near characteristic
interband transition energies, with enhanced shifts in the
vicinity of the pseudo-Brewster angle. Our results connect
altermagnetic band topology, optical Hall response, and
helicity-dependent beam shifts, suggesting an optical approach
to probing Chern-insulating phases and their interband spectra.
\endgroup

The remainder of this paper is organized as follows.
Section~\ref{sec:model} introduces the lattice Hamiltonian
and topological properties of the altermagnetic system.
Section~\ref{sec:conductivity} presents the optical
conductivity formalism based on the Kubo approach.
Section~\ref{sec:fresnel} develops the generalized Fresnel
reflection coefficients and the photonic spin Hall
shift formalism. Section~\ref{sec:results} presents
and discusses the numerical results, while
Sec.~\ref{sec:conclusions} summarizes the main findings.

\section{Model, topology, and symmetries}
\label{sec:model}

\subsection{Hamiltonian and bands}
\label{sec:hamiltonian}
The interplay between Rashba spin--orbit coupling,
momentum-dependent $d$-wave altermagnetic exchange,
and an out-of-plane Zeeman field has been investigated
within a continuum framework by Rao
\textit{et al.}~\cite{Rao2024}.
Here, we adopt the corresponding two-band lattice
formulation considered by Qin
\textit{et al.}~\cite{Qin2026}, which retains the
full Brillouin-zone structure and allows us to
examine band inversions and Chern-insulating
phases at the high-symmetry points.
The Hamiltonian is written as
\begin{equation}
\hat{H}(\mathbf{k})
=\hat{H}_{\mathrm{soc}}(\mathbf{k})
+\hat{H}_{J}(\mathbf{k})
+\hat{H}_{z},
\label{eq:H}
\end{equation}
where
\begin{align}
\hat{H}_{\mathrm{soc}}(\mathbf{k})
&=t\sin(k_ya)\sigma_x
-t\sin(k_xa)\sigma_y,
\label{eq:Hsoc}\\
\hat{H}_{J}(\mathbf{k})
&=2t_J[\cos(k_xa)-\cos(k_ya)]\sigma_z,
\label{eq:HJ}\\
\hat{H}_{z}
&=\Delta_z\sigma_z.
\label{eq:Hz}
\end{align}
Here $\sigma_{x,y,z}$ are Pauli matrices acting on spin and $a$ is the lattice constant. The first term is a Rashba spin--orbit coupling of strength $t$, the second is the $d$-wave altermagnetic term of strength $t_J$, and the third is an exchange (Zeeman) field $\Delta_z$ normal to the sheet. The third contribution acts as a uniform spin splitting and must not be conflated with the vanishing-net-moment altermagnetic order itself. Writing $\hat{H}=\mathbf{d}(\mathbf{k})\cdot\boldsymbol{\sigma}$ with
\begin{align}
\mathbf{d}(\mathbf{k}) &= \big(t\sin k_y a,\; -t\sin k_x a,\; d_z\big), \nonumber\\
d_z &= 2t_J\,(\cos k_x a-\cos k_y a)+\Delta_z, \label{eq:dvec}
\end{align}
the two bands are $\varepsilon_{\pm}(\mathbf{k})=\pm\varepsilon_{\mathbf{k}}$, with
\begin{equation}
\varepsilon_{\mathbf{k}}=|\mathbf{d}(\mathbf{k})|=\sqrt{t^{2}\sin^{2}k_xa+t^{2}\sin^{2}k_ya+d_z^{2}}.
\label{eq:bands}
\end{equation}
We use $t=1$~eV, $a=1$, $E_F=0$, and zero temperature throughout. Unless otherwise stated, the reference calculations take $t_J=0.6$~eV and $\Delta_z=-0.5$~eV; the phase diagram varies $t_J$ and $\Delta_z$ independently, while other explicitly stated configurations are used for the band and optical comparisons.

\subsection{Gap closings and Chern number}
\label{sec:chern}

At the four high-symmetry points $\Gamma=(0,0)$, $\mathrm{M}=(\pi/a,\pi/a)$, $\mathrm{X}=(\pi/a,0)$, and $\mathrm{Y}=(0,\pi/a)$ the Rashba term vanishes, and the gap reduces to $2|d_z|$. Direct substitution into Eq.~\eqref{eq:dvec} gives
\begin{equation}
\begin{aligned}
d_z(\Gamma)&=d_z(\mathrm{M})=\Delta_z,\\
d_z(\mathrm{X})&=\Delta_z-4t_J,\\
d_z(\mathrm{Y})&=\Delta_z+4t_J,
\end{aligned}
\label{eq:masses}
\end{equation}
so the gap closes at $\Delta_z=0$ and at $\Delta_z=\pm4t_J$. These X and Y labels follow from direct substitution. Near each of the four points the Hamiltonian reduces to a massive Dirac Hamiltonian; $\Gamma$ and M have one chirality, and X and Y the opposite one. Summing the half-integer contributions of the four gapped Dirac points gives the Chern number of the lower band~\cite{Cheraghchi2026,Qin2026},
\begin{equation}
C=\tfrac{1}{2}\left[\operatorname{sgn}(\Delta_z+4t_J)+\operatorname{sgn}(\Delta_z-4t_J)-2\operatorname{sgn}(\Delta_z)\right],
\label{eq:C}
\end{equation}
For fixed $t_J\neq0$, Eq.~\eqref{eq:C} gives $C=-\operatorname{sgn}(\Delta_z)$ for $|\Delta_z|<4|t_J|$ and $C=0$ otherwise; $C=0$ when $t_J=0$.

The same result follows from the Berry curvature. For a two-band Hamiltonian $\mathbf{d}\cdot\boldsymbol{\sigma}$, the Berry curvature of band $s=\pm$ is
\begin{equation}
\Omega_{s}(\mathbf{k})=-\frac{s}{2}\,
\frac{\mathbf{d}\cdot\left(\partial_{k_x}\mathbf{d}\times\partial_{k_y}\mathbf{d}\right)}{\varepsilon_{\mathbf{k}}^{3}},
\label{eq:Omega_general}
\end{equation}
which for the present model becomes
\begin{align}
\Omega_\pm(\mathbf{k}) &= \pm\frac{a^2t^2J_u(\mathbf{k})}{2\varepsilon_{\mathbf{k}}^3}, \nonumber\\
J_u(\mathbf{k}) &= 2t_J\,(\cos k_x a-\cos k_y a)-\Delta_z\cos k_x a\,\cos k_y a. \label{eq:berry}
\end{align}
The Chern number of the lower band is then
\begin{equation}
C=-\int_{\mathrm{BZ}}\frac{d^{2}k}{2\pi}\,\Omega_{-}(\mathbf{k}).
\label{eq:C_numeric}
\end{equation}
The minus sign in Eq.~\eqref{eq:C_numeric} specifies the orientation convention adopted for reporting $C$ and its Hall-conductivity sign; comparisons with other Berry-curvature conventions require consistent band and momentum-space orientations.


\section{Optical conductivity}
\label{sec:conductivity}
The frequency-dependent optical conductivity tensor is
evaluated using the Kubo formalism \cite{Shah2024},
\begin{widetext}
\begin{equation}
\sigma_{\alpha\beta}(\omega)
=
-i\frac{e^2}{h}
\sum_{m,n}
\int_{\mathrm{BZ}}\frac{d^2k}{2\pi}
\frac{
\langle m|\partial_\alpha\hat{H}|n\rangle
\langle n|\partial_\beta\hat{H}|m\rangle
}{
\varepsilon_{m\mathbf{k}}-\varepsilon_{n\mathbf{k}}
+\hbar\omega+i\eta
}
\frac{
\Theta(E_F-\varepsilon_{m\mathbf{k}})
-\Theta(E_F-\varepsilon_{n\mathbf{k}})
}{
\varepsilon_{m\mathbf{k}}-\varepsilon_{n\mathbf{k}}
},
\label{eq:optical_conductivity}
\end{equation}
\end{widetext}
Here, $\alpha,\beta\in\{x,y\}$ denotes Cartesian
components, and
$\partial_\alpha\equiv\partial/\partial k_\alpha$
represents differentiation with respect to the
corresponding crystal momentum. The states
$|m\rangle$ and $|n\rangle$ are eigenstates of
the Hamiltonian $\hat{H}(\mathbf{k})$, with
eigenenergies $\varepsilon_{m\mathbf{k}}$ and
$\varepsilon_{n\mathbf{k}}$, respectively.
Furthermore, $-e$ denotes the electron charge,
$E_F$ is the Fermi energy, $\eta>0$ is the
phenomenological broadening parameter, and
$\Theta(x)$ represents the Heaviside step
function. The summation extends over pairs
of electronic states connected by optically
allowed interband transitions.

\begingroup\color{black}
\section{Reflection from a Hall sheet}
\label{sec:fresnel}

\subsection{Reflection coefficients and electromagnetic
boundary conditions}
\label{sec:reflection}
\begin{figure}[t]
    \centering
    \includegraphics[width=\columnwidth]{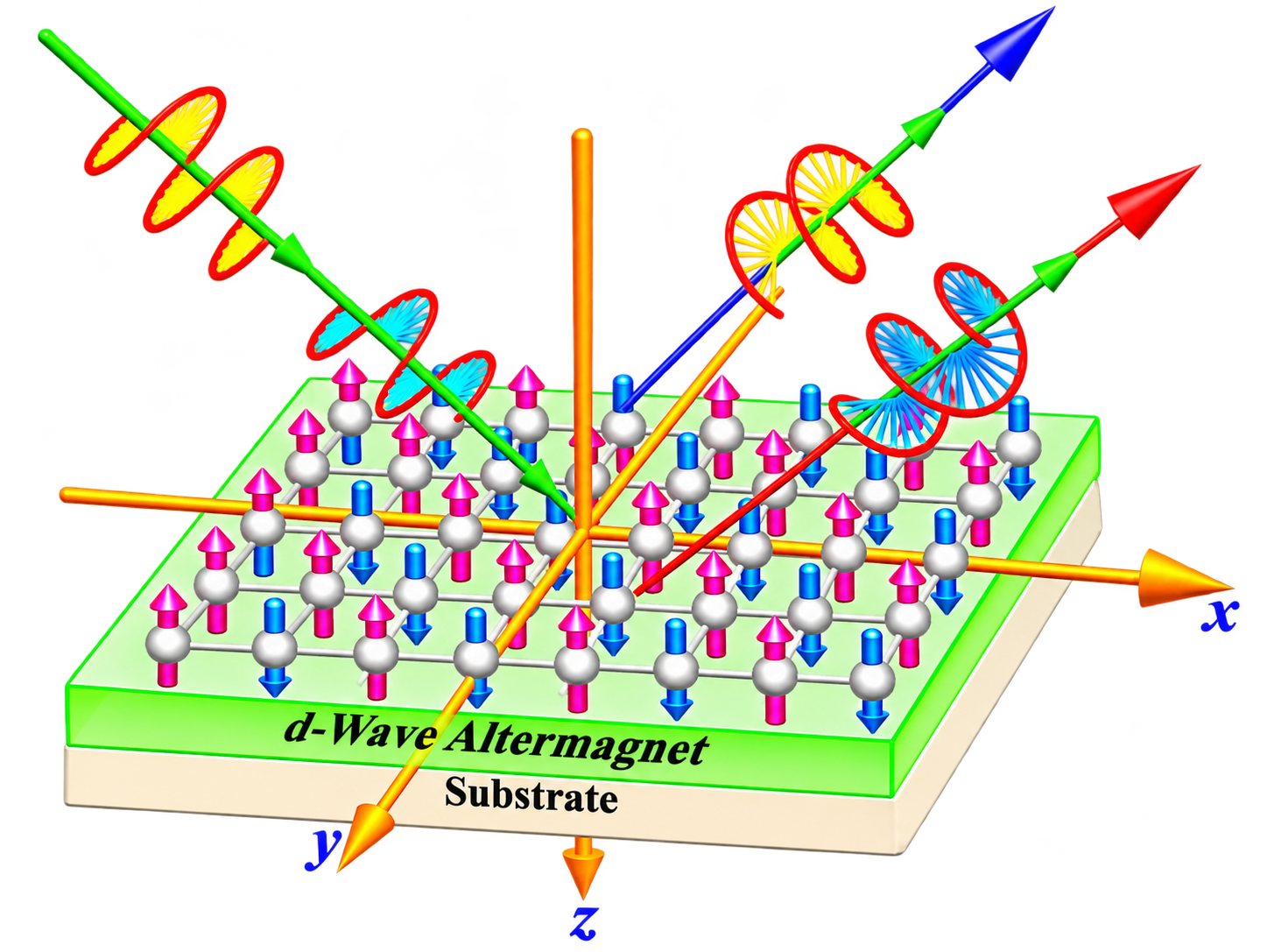}
    \caption{Schematic illustration of light reflection at the surface
    of a $d$-wave altermagnet in a Cartesian coordinate system.
    The altermagnetic layer, characterized by two oppositely
    aligned magnetic sublattices, is placed on a homogeneous
    and isotropic substrate. Upon reflection, the photonic
    spin Hall effect (PSHE) produces spin-dependent spatial
    splitting of the reflected beam, giving rise to both
    in-plane (longitudinal) and transverse spin Hall shifts.}
    \label{fig:PSHE_altermagnet}
\end{figure}
We consider a two-dimensional conducting sheet located at
$z=0$, separating air ($n_i=1$, $z>0$) from a dielectric
substrate of refractive index $n_t$ ($z<0$) [see Fig. \ref{fig:PSHE_altermagnet}]. The conducting
layer is treated within the zero-thickness approximation,
and its electromagnetic response is characterized by the
frequency-dependent optical conductivity tensor
$\hat{\sigma}(\omega)$. A monochromatic plane wave is incident from the air side
at an angle $\theta_i$ with respect to the surface normal,
with the plane of incidence chosen as the $x$--$z$ plane.
The transmission angle $\theta_t$ is determined by
Snell's law, $n_i\sin\theta_i=n_t\sin\theta_t$, yielding
\begin{equation}
\cos\theta_t=
\sqrt{1-\left(\frac{n_i}{n_t}
\sin\theta_i\right)^2}.
\label{eq:snell}
\end{equation}
To describe the polarization-dependent reflection,
we introduce the orthogonal polarization basis
$\hat{\mathbf e}_s=\hat{\mathbf y}$ and
$\hat{\mathbf e}_p=
\hat{\mathbf k}\times\hat{\mathbf e}_s$.
For signed transverse displacements in the beam-shift section, we use the laboratory-positive reflected axis $y_r=+y$ and circular analyzers $E_{r\pm}=(E_r^p\pm iE_r^s)/\sqrt{2}$, as in the accompanying Maxwell implementation. With this convention, the incident and reflected
$p$-polarization unit vectors are given by
\begin{equation}
\begin{aligned}
\hat{\mathbf e}_{pi}
&=(\cos\theta_i,0,\sin\theta_i),\\
\hat{\mathbf e}_{pr}
&=(-\cos\theta_i,0,\sin\theta_i).
\end{aligned}
\label{eq:basis}
\end{equation}
The reflected electric-field amplitudes are related to
the incident amplitudes through the reflection matrix
\begin{equation}
\begin{pmatrix}
E_r^p\\
E_r^s
\end{pmatrix}
=
\begin{pmatrix}
r_{pp}&r_{ps}\\
r_{sp}&r_{ss}
\end{pmatrix}
\begin{pmatrix}
E_i^p\\
E_i^s
\end{pmatrix},
\label{eq:Rmatrix}
\end{equation}
where the first and second indices denote the reflected
and incident polarization components, respectively.
The diagonal coefficients $r_{pp}$ and $r_{ss}$
describe polarization-preserving reflection, whereas
$r_{ps}$ and $r_{sp}$ characterize cross-polarization
conversion induced by the off-diagonal components
of the optical conductivity tensor. The electromagnetic response of the conducting sheet
is determined by the boundary conditions at $z=0$.
The tangential component of the electric field is
continuous across the interface, whereas the
tangential magnetic field exhibits a discontinuity
associated with the induced surface current.
The boundary conditions can be expressed as
\begin{subequations}
\label{eq:boundary}
\begin{align}
\mathbf E_{\parallel}^{\mathrm{above}}
&=
\mathbf E_{\parallel}^{\mathrm{below}},
\\
\hat{\mathbf z}\times
\left(
\mathbf H_{\mathrm{above}}
-
\mathbf H_{\mathrm{below}}
\right)
&=
\mathbf J_s,
\end{align}
\end{subequations}
where
\begin{equation}
\mathbf J_s=
\hat{\sigma}_{\mathrm{phys}}
\mathbf E_{\parallel}
\label{eq:surfacecurrent}
\end{equation}
is the induced sheet current density.
Here, $\hat{\sigma}_{\mathrm{phys}}$ denotes the
physical two-dimensional conductivity tensor.

Introducing the dimensionless conductivity tensor
$\hat{\sigma}=Z_0\hat{\sigma}_{\mathrm{phys}}$,
where $Z_0$ is the vacuum impedance, and expressing
the magnetic fields in units of $1/Z_0$,
the boundary equations can be solved analytically
to obtain the generalized Fresnel reflection
coefficients:
\begin{subequations}
\label{eq:fresnel}
\begin{align}
r_{pp}
&=
\frac{
\alpha_-\beta_+
-\sigma_{xy}\sigma_{yx}
}{D},
\label{eq:fresnel_rpp}
\\
r_{ss}
&=
-\frac{
\alpha_+\beta_-
-\sigma_{xy}\sigma_{yx}
}{D},
\label{eq:fresnel_rss}
\\
r_{ps}
&=
\frac{2n_i\sigma_{xy}}{D},
\label{eq:fresnel_rps}
\\
r_{sp}
&=
-\frac{2n_i\sigma_{yx}}{D}.
\label{eq:fresnel_rsp}
\end{align}
\end{subequations}
Here, the common denominator is defined as
\begin{equation}
D=
\alpha_+\beta_+
-\sigma_{xy}\sigma_{yx},
\label{eq:denominator}
\end{equation}
with
\begin{subequations}
\label{eq:alphabeta}
\begin{align}
\alpha_\pm
&=
\sigma_{xx}\cos\theta_i
+
\frac{n_t\cos\theta_i}{\cos\theta_t}
\pm n_i,
\\
\beta_\pm
&=
\frac{\sigma_{yy}}{\cos\theta_i}
+
\frac{n_t\cos\theta_t}{\cos\theta_i}
\pm n_i.
\end{align}
\end{subequations}
Equations~\eqref{eq:fresnel}--\eqref{eq:alphabeta}
describe the reflection response of a general
two-dimensional conducting interface, including
anisotropic and nonreciprocal optical conductivity
tensors. For the $d$-wave altermagnetic system considered
here, the optical conductivity satisfies
\begin{equation}
\sigma_{yy}=\sigma_{xx},
\qquad
\sigma_{yx}=-\sigma_{xy},
\label{eq:isotropy}
\end{equation}
so that the denominator becomes
\begin{equation}
D=
\alpha_+\beta_++\sigma_{xy}^2.
\label{eq:hall_denominator}
\end{equation}
Consequently, the reflection coefficients obey
the following parity relations under reversal
of the Hall conductivity:
\begin{equation}
\begin{aligned}
r_{pp}(-\sigma_{xy})
&=r_{pp}(\sigma_{xy}),
\\
r_{ss}(-\sigma_{xy})
&=r_{ss}(\sigma_{xy}),
\\
r_{ps}(-\sigma_{xy})
&=-r_{ps}(\sigma_{xy}),
\\
r_{sp}(-\sigma_{xy})
&=-r_{sp}(\sigma_{xy}).
\end{aligned}
\label{eq:parity_r}
\end{equation}
Moreover, the cross-polarized reflection
coefficients become identical,
\begin{equation}
r_{ps}=r_{sp}
=
\frac{2n_i\sigma_{xy}}{D}.
\label{eq:hall_cross}
\end{equation}
An immediate consequence of this relation is that
the combination
$\rho\equiv r_{ps}-r_{sp}$
vanishes identically for the antisymmetric
Hall conductivity tensor. In contrast,
a symmetric off-diagonal conductivity tensor
with $\sigma_{yx}=\sigma_{xy}$ yields
$r_{ps}=-r_{sp}$. This distinction provides
a useful means of identifying the symmetry
of the optical conductivity through
polarization-dependent reflection measurements.

A salient aspect of the reflection response is the existence of a pronounced minimum in the $p$-polarized reflection amplitude in the vicinity of the Brewster angle for an uncoated dielectric interface,
\begin{equation}
\theta_B=
\tan^{-1}\left(\frac{n_t}{n_i}\right).
\label{eq:brewster}
\end{equation}
For a conducting interface, the nonzero and generally complex optical conductivity alters the standard Brewster condition, giving rise to a shifted pseudo-Brewster minimum. In the vicinity of this minimum, the diagonal reflection coefficient $r_{pp}$ is strongly suppressed, whereas the Hall-induced cross-polarized coefficient $r_{sp}\propto\sigma_{xy}$ may remain finite. Consequently, the relative weight of the cross-polarized component in the reflected field increases, yielding a marked polarization-dependent response. This selective suppression enhances the helicity-dependent beam displacement, rendering the pseudo-Brewster regime a particularly suitable parameter range for probing the anomalous Hall contribution to the optical response and, by extension, the underlying topological characteristics of the $d$-wave altermagnet.

\endgroup

\begingroup\color{black}
\subsection{Photonic spin Hall effect}
\label{sec:shifts}
Having obtained the longitudinal and transverse optical conductivities, together with the associated Fresnel reflection coefficients, we proceed to analyze the photonic spin Hall effect (PSHE) for a Gaussian beam reflected from the surface of a two-dimensional $d$-wave altermagnet. The PSHE arises from the spin–orbit interaction of light, which induces coupling between the polarization and spatial degrees of freedom and thereby generates helicity-dependent transverse and longitudinal displacements of the reflected beam. Using the angular-spectrum approach~\cite{BliokhAiello2013}, we compute the spin-dependent beam shifts by incorporating the optical response functions of the altermagnetic system. Specifically, we consider a horizontally ($H$) or vertically ($V$) polarized Gaussian beam incident at an angle $\theta_i$ onto an altermagnetic sheet supported by a dielectric substrate. The angular spectrum of the incident beam can be expressed as
\begin{equation}
G(\kappa_x,\kappa_y)
=
G_0
\exp\left[
-\frac{w_0^2}{4}
\left(\kappa_x^2+\kappa_y^2\right)
\right],
\label{eq:gaussian_pshe}
\end{equation}
where $w_0$ is the beam waist, $\kappa_x$ and
$\kappa_y$ denote the transverse wave-vector
components in the reflected-beam frame, and
$G_0$ is a normalization constant.

The reflected electric field can be expressed as the superposition of two circularly polarized components, indexed by the helicity $\pm$. Adopting the specified polarization convention, we define
\begin{equation}
E_{r\pm}
=
\frac{1}{\sqrt{2}}
\left(E_r^p\pm iE_r^s\right).
\label{eq:pshe_circular}
\end{equation}
The corresponding helicity-resolved reflected amplitudes are governed by the generalized Fresnel reflection coefficients given in Eq.~\eqref{eq:fresnel}. For a paraxial incident beam, each plane-wave constituent is characterized by a slightly perturbed angle of incidence and an associated plane of incidence. A first-order expansion of the reflection response with respect to the transverse wave-vector components yields
\begin{equation}
\widetilde E_{r\pm}^{\,\nu}
=
G(\kappa_x,\kappa_y)
\left[
A_{0\pm}^{\nu}
+
\frac{\kappa_x}{k_0}A_{x\pm}^{\nu}
+
\frac{\kappa_y}{k_0}A_{y\pm}^{\nu}
\right],
\label{eq:pshe_angular}
\end{equation}
where $\nu=H,V$, $k_0=\omega/c$, and the coefficients $A_{0\pm}^{\nu}$, $A_{x\pm}^{\nu}$, and $A_{y\pm}^{\nu}$ respectively represent the central reflected amplitude and its first-order angular dependence. For a horizontally polarized incident field, these coefficients take the form
\begin{subequations}
\label{eq:pshe_amplitudes_H}
\begin{align}
A_{0\pm}^{H}
&=r_{pp}\pm i r_{sp},
\\
A_{x\pm}^{H}
&=-\left(r_{pp}'\pm i r_{sp}'\right),
\\
A_{y\pm}^{H}
&=\mp i\Phi\cot\theta_i,
\end{align}
\end{subequations}
where $\Phi=r_{pp}+r_{ss}$ and $r_{ab}'=\partial r_{ab}/\partial\theta_i$. Unless otherwise stated, all reflection coefficients and their angular derivatives are evaluated at the central incidence angle. For vertically polarized incident light, the corresponding leading-order amplitudes depend on the reflected-axis convention and are stated here for formal comparison; their full-vector numerical validation is not included in the present $H$-input figures. Similarly, we obtain
\begin{subequations}
\label{eq:pshe_amplitudes_V}
\begin{align}
A_{0\pm}^{V}
&=r_{sp}\pm i r_{ss},
\\
A_{x\pm}^{V}
&=-\left(r_{sp}'\pm i r_{ss}'\right),
\\
A_{y\pm}^{V}
&=-\Phi\cot\theta_i.
\end{align}
\end{subequations}
Here, we have used the Hall-sheet symmetry
$r_{ps}=r_{sp}$, which follows from
$\sigma_{yx}=-\sigma_{xy}$ under our
polarization convention. Consequently,
the combination
$\rho=r_{ps}-r_{sp}$ vanishes, simplifying
the transverse angular response. The spatial displacement of each circular
component is determined from the centroid
of its reflected intensity distribution,
\begin{equation}
\Delta u_{r\pm}^{\nu}
=
\frac{
\displaystyle
\iint u_r
\left|E_{r\pm}^{\nu}(x_r,y_r)\right|^2
\,dx_r\,dy_r
}{
\displaystyle
\iint
\left|E_{r\pm}^{\nu}(x_r,y_r)\right|^2
\,dx_r\,dy_r
},
\label{eq:pshe_centroid}
\end{equation}
where $u=x,y$ and $\nu=H,V$.
The longitudinal displacement corresponds
to $u=x$, whereas the transverse
displacement corresponds to $u=y$. We characterize the spin-dependent
separation of the reflected circular
components by
\begin{subequations}
\label{eq:pshe_spin_definitions}
\begin{align}
S_x^\nu
&=
\frac{
\Delta x_{r+}^{\nu}
-
\Delta x_{r-}^{\nu}
}{2},
\\
S_y^\nu
&=
\frac{
\Delta y_{r+}^{\nu}
-
\Delta y_{r-}^{\nu}
}{2}.
\end{align}
\end{subequations}
Thus, $S_x^\nu$ and $S_y^\nu$ represent
half of the relative longitudinal and
transverse separations of the two
circularly polarized components,
respectively. For the first-order angular spectrum
in Eq.~\eqref{eq:pshe_angular}, direct
evaluation of the intensity centroid
gives
\begin{equation}
\Delta u_{r\pm}^{\nu}
=
-\frac{1}{k_0}
\frac{
\mathrm{Im}\left[
\left(A_{0\pm}^{\nu}\right)^*
A_{u\pm}^{\nu}
\right]
}{
\displaystyle
\left|A_{0\pm}^{\nu}\right|^2
+
\frac{
\left|A_{x\pm}^{\nu}\right|^2+
\left|A_{y\pm}^{\nu}\right|^2
}{
(k_0w_0)^2
}
},
\label{eq:pshe_finite}
\end{equation}
where the denominator accounts for the
finite angular width of the Gaussian beam.
This expression is exact for the reflected
spectrum truncated at first order in
$\kappa_x$ and $\kappa_y$. Away from reflection minima, where
\begin{equation}
\frac{
|A_{x\pm}^{\nu}|^2+
|A_{y\pm}^{\nu}|^2
}{
(k_0w_0)^2
}
\ll
|A_{0\pm}^{\nu}|^2,
\label{eq:pshe_weak_condition}
\end{equation}
the leading-order displacement reduces to
\begin{equation}
\Delta u_{r\pm}^{\nu}
\simeq
-\frac{1}{k_0}
\mathrm{Im}
\left(
\frac{A_{u\pm}^{\nu}}
{A_{0\pm}^{\nu}}
\right).
\label{eq:pshe_firstorder}
\end{equation}
Substituting Eqs.~\eqref{eq:pshe_amplitudes_H}
and \eqref{eq:pshe_amplitudes_V} into
Eq.~\eqref{eq:pshe_firstorder}, and then
using Eq.~\eqref{eq:pshe_spin_definitions},
we obtain closed expressions for the
spin-dependent longitudinal and transverse
displacements. For horizontally polarized incident light,
the longitudinal spin-dependent shift is
\begin{equation}
S_x^H
=
\frac{1}{k_0}
\mathrm{Re}
\left[
\frac{
r_{pp}\,\partial_{\theta_i}r_{sp}
-
r_{sp}\,\partial_{\theta_i}r_{pp}
}{
r_{pp}^{\,2}+r_{sp}^{\,2}
}
\right].
\label{eq:SxH}
\end{equation}
The corresponding transverse
spin-dependent shift becomes
\begin{equation}
S_y^H
=
+\frac{\cot\theta_i}{k_0}
\mathrm{Re}
\left[
\frac{
(r_{pp}+r_{ss})r_{pp}
}{
r_{pp}^{\,2}+r_{sp}^{\,2}
}
\right].
\label{eq:SyH}
\end{equation}
For the indicated $V$-polarization angular-spectrum amplitudes, the corresponding leading-order expression is retained for formal comparison (no $V$-input numerical curves are reported). Similarly, the longitudinal shift
is given by
\begin{equation}
S_x^V
=
\frac{1}{k_0}
\mathrm{Re}
\left[
\frac{
r_{sp}\,\partial_{\theta_i}r_{ss}
-
r_{ss}\,\partial_{\theta_i}r_{sp}
}{
r_{ss}^{\,2}+r_{sp}^{\,2}
}
\right].
\label{eq:SxV}
\end{equation}
The corresponding transverse shift reads
\begin{equation}
S_y^V
=
-\frac{\cot\theta_i}{k_0}
\mathrm{Re}
\left[
\frac{
(r_{pp}+r_{ss})r_{ss}
}{
r_{ss}^{\,2}+r_{sp}^{\,2}
}
\right].
\label{eq:SyV}
\end{equation}
Equations~\eqref{eq:SxH}--\eqref{eq:SyV}
constitute the leading-order expressions
for the longitudinal and transverse
spin-dependent separations in the
altermagnetic system. They explicitly
connect the PSHE to the complex Fresnel
reflection coefficients and, consequently,
to the longitudinal and transverse
optical conductivities of the material. For the two topologically nontrivial
altermagnetic phases with opposite Chern
numbers, the Hall conductivity changes
sign, whereas the longitudinal
conductivity remains approximately
unchanged. As a consequence, the
cross-polarized reflection amplitude
$r_{sp}$ is odd under Hall reversal,
while the diagonal coefficients
$r_{pp}$ and $r_{ss}$ are even. Within the leading-order expressions,
these symmetry properties imply that
the longitudinal spin-dependent shifts
$S_x^H$ and $S_x^V$ are odd under the reversal
of the Hall conductivity, whereas the
transverse shifts $S_y^H$ and $S_y^V$
are even. This distinction provides
a useful optical signature for probing
the sign of the anomalous Hall response
in topologically nontrivial altermagnets.

\endgroup

\section{Results and discussion}
\label{sec:results}

\subsection{Chern phase diagram and band structures}
\begin{figure}[t]
\centering
\includegraphics[width=1.0\columnwidth]{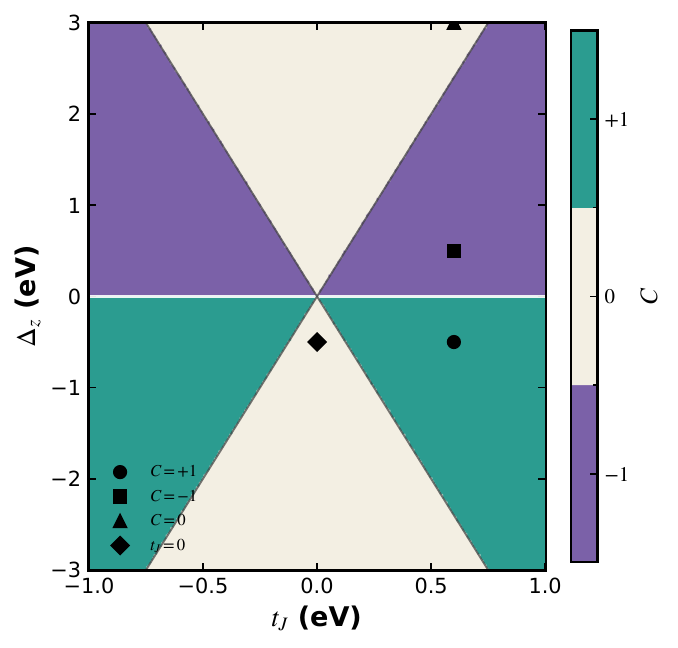}
\caption{\textcolor{black}{Chern phase diagram in the $(t_J,\Delta_z)$ plane for $t=1$~eV, obtained by independently scanning the altermagnetic exchange $t_J$ and uniform exchange $\Delta_z$. The markers identify the band-structure points at $t_J=0.6$~eV and the separate $t_J=0$ optical control. The gap closes at $\Delta_z=0,\pm4t_J$; the Chern number is undefined on the gap-closing boundaries.}}
\label{fig:phase}
\end{figure}

\begin{figure}[t]
\includegraphics[width=\columnwidth]{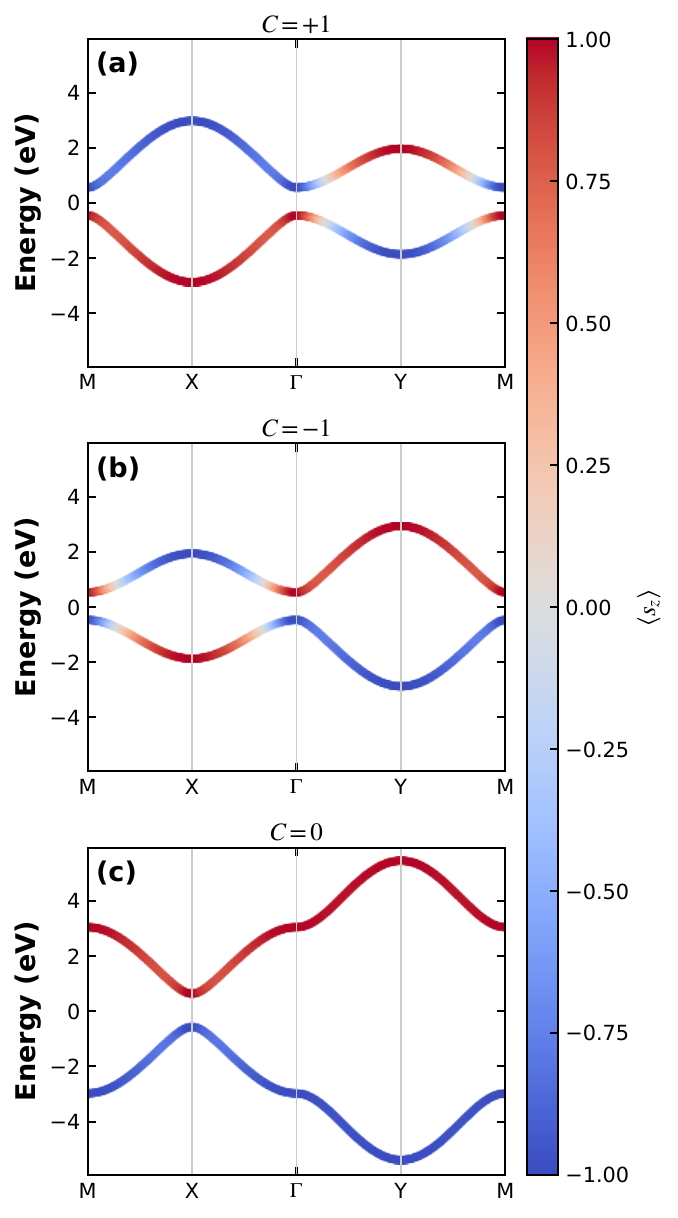}
\caption{
Band structures of the $d$-wave altermagnetic model
along the M--X--$\Gamma$--Y--M high-symmetry path
for three distinct topological phases:
(a) the Chern-insulating phase with $C=+1$ at
$\Delta_z=-0.5$~eV,
(b) the Chern-insulating phase with $C=-1$ at
$\Delta_z=+0.5$~eV, and
(c) the topologically trivial phase with $C=0$ at
$\Delta_z=+3.0$~eV.
The altermagnetic exchange parameter is fixed at
$t_J=0.6$~eV in all three panels.
The color scale represents the spin polarization
$\langle s_z\rangle$ of the corresponding
electronic bands.
}
\label{fig:bands}
\end{figure}

\begingroup\color{black}
Throughout, we set $t=1$~eV, $a=1$, $E_F=0$, and assume zero temperature. Figure~\ref{fig:phase} first presents the Chern-number phase diagram obtained by independently tuning $t_J$ and $\Delta_z$; accordingly, $t_J$ is not held fixed in this panel. In the subsequent calculations of representative band structures and optical response functions, we fix $t_J=0.6$~eV unless an explicit $t_J=0$ control case is specified. The phases with $C=0$ and $C=\pm1$ are separated by band-gap closures at $\Delta_z=0$ and $\Delta_z=\pm4t_J$, consistent with Eq.~\eqref{eq:C}. For $t_J=0.6$~eV, numerical evaluation of Eq.~\eqref{eq:C_numeric} yields $C=+1.000000$ at $\Delta_z=-0.5$~eV, $C=-1.000000$ at $\Delta_z=+0.5$~eV, and $C=0.000000$ at $\Delta_z=-2.8$~eV, in agreement with the analytic phase boundaries.

Figure~\ref{fig:bands}(a)--\ref{fig:bands}(c) subsequently compares the electronic band structures along M--X--$\Gamma$--Y--M for representative $C=+1$, $C=-1$, and $C=0$ insulating phases, respectively. The corresponding exchange fields are $\Delta_z=-0.5$, $+0.5$, and $+3.0$~eV at fixed $t_J=0.6$~eV. For $\Delta_z=0$, the gap closes at $\Gamma$ and M, while nonzero uniform exchange opens a direct gap. For the displayed $C=\pm1$ configurations, the minimum direct gap is $2|\Delta_z|$. The color scale represents the spin polarization $\langle s_z\rangle=\pm d_z/\varepsilon_{\mathbf{k}}$. The two opposite-Chern phases have equal-sized gaps at the reference exchange fields but different spin-polarization patterns near the inequivalent X and Y extrema. The trivial case lies outside the $|\Delta_z|=4|t_J|$ band-inversion region. This $C=0$ band-structure configuration ($t_J=0.6$~eV, $\Delta_z=3.0$~eV) differs from the $t_J=0$ trivial control included in the optical-conductivity plots.
\endgroup

\begin{figure*}[ht!]
\centering
\includegraphics[width=\textwidth]{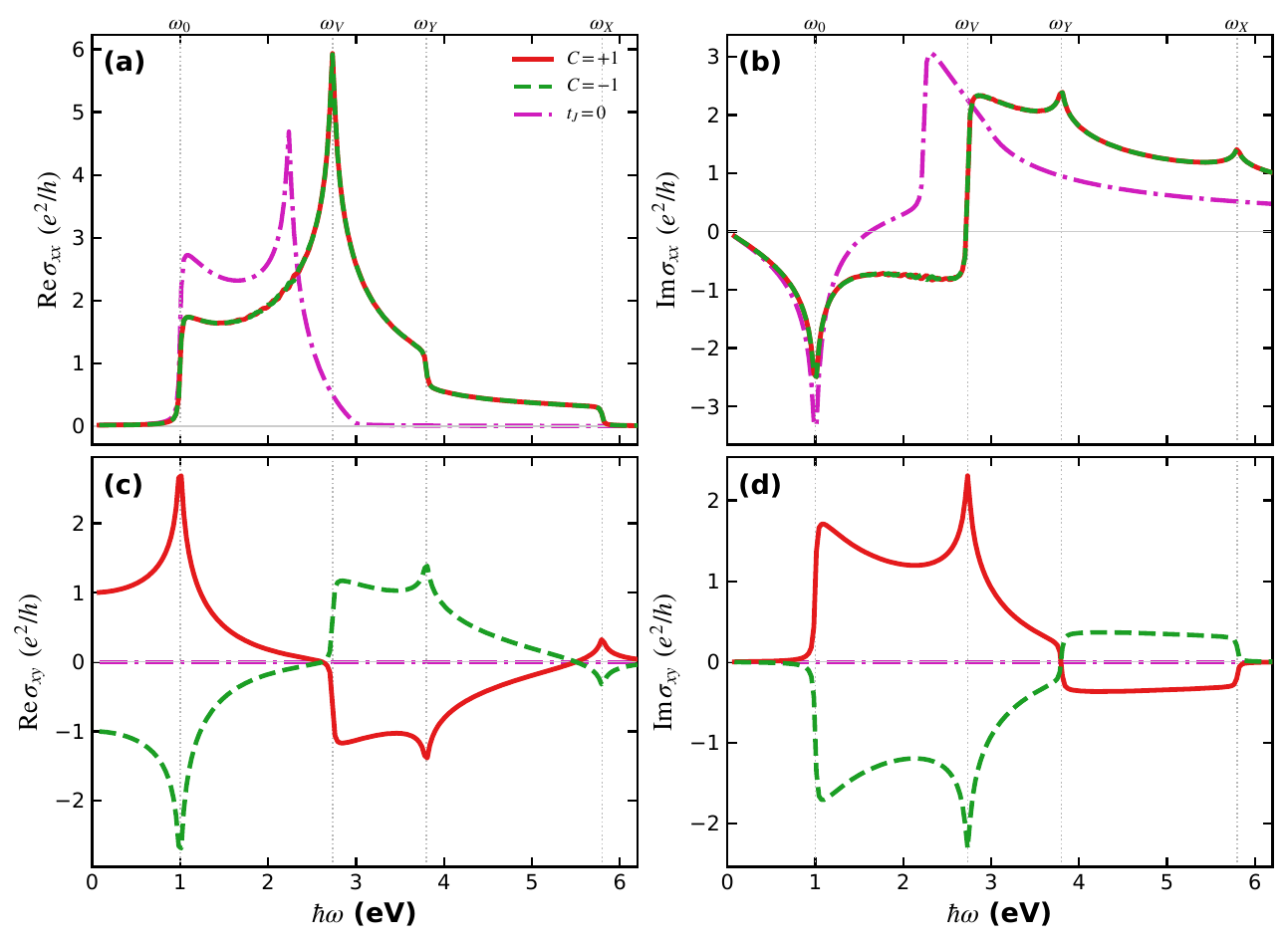}
\caption{Optical conductivity of the altermagnetic sheet for $t_J=0.6$~eV in the $C=+1$ phase ($\Delta_z=-0.5$~eV, red solid) and the $C=-1$ phase ($\Delta_z=+0.5$~eV, green dashed), together with the $t_J=0$ control case ($\Delta_z=-0.5$~eV, magenta dash-dotted), in units of $e^2/h$. (a) Re\,$\sigma_{xx}$, (b) Im\,$\sigma_{xx}$, (c) Re\,$\sigma_{xy}$, and (d) Im\,$\sigma_{xy}$. Dashed and dash-dotted styles are retained where curves overlap to keep the separate cases visible.}
\label{fig:conductivity}
\end{figure*}

\subsection{Characteristic optical energies}
\label{sec:energies}
\begingroup\color{black}
The relevant optical thresholds follow from the direct interband gap $2\varepsilon_{\mathbf{k}}$. At the reference parameters $t=1$~eV, $t_J=0.6$~eV, and $\Delta_z=-0.5$~eV, the minimum-gap transition occurs at $\hbar\omega_0=2|\Delta_z|=1.00$~eV. The energy associated with an off-axis saddle point of the reference dispersion is
\begin{equation}
\hbar\omega_V=2|t|\sqrt{2+\frac{\Delta_z^2}{t^2-8t_J^2}}\simeq2.733~\mathrm{eV},
\label{eq:omegaV}
\end{equation}
provided the stationary-point solution exists in the Brillouin zone. The inequivalent high-symmetry-point gaps are $\hbar\omega_Y=2|\Delta_z+4t_J|=3.80$~eV and $\hbar\omega_X=2|\Delta_z-4t_J|=5.80$~eV. 
\endgroup
\begin{figure*}[ht!]
\centering
\includegraphics[width=\textwidth]{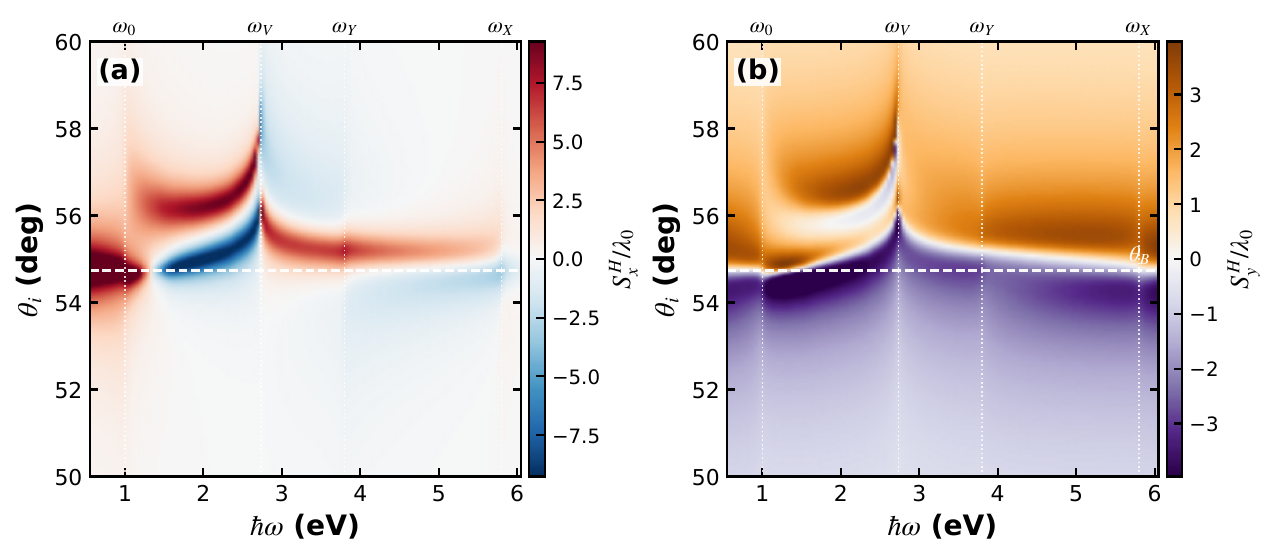}
\caption{\textcolor{black}{Finite-beam spin splittings $S_x^H/\lambda_0$ (a) and $S_y^H/\lambda_0$ (b) as functions of photon energy and incidence angle for $C=+1$. White dotted markers indicate the characteristic energies $\hbar\omega_0=1.00$~eV, $\hbar\omega_V\simeq2.73$~eV, $\hbar\omega_{Y}=3.80$~eV and $\hbar\omega_X=5.80$~eV, while the horizontal dashed line marks the bare-interface Brewster angle. Data use $t=1$~eV, $t_J=0.6$~eV, $\Delta_z=-0.5$~eV, $\eta=0.015$~eV, $n_t=\sqrt{2}$, and $w_0=50\lambda_0$. The color limits are individually clipped for visualization; numerical values are unchanged.}}
\label{fig:shifts}
\end{figure*}
\subsection{Optical conductivities and spectral signatures}
Figure~\ref{fig:conductivity}(a)--(d) displays the
real and imaginary parts of the longitudinal
$\sigma_{xx}(\omega)$ and transverse
$\sigma_{xy}(\omega)$ optical conductivities
for three representative parameter configurations.
The red solid and green dashed curves correspond
to the Chern-insulating phases $C=+1$
($\Delta_z=-0.5$~eV) and $C=-1$
($\Delta_z=+0.5$~eV), respectively, with
$t_J=0.6$~eV, whereas the magenta dash-dotted
curves represent the trivial reference system
with $t_J=0$ and $\Delta_z=-0.5$~eV. Figure~\ref{fig:conductivity}(a) shows the real
part of the longitudinal conductivity as a function of photon energy $\hbar\omega$.
For photon energies below the minimum interband
excitation threshold $\hbar\omega_0=1.00$~eV,
the conductivity is strongly suppressed,
reflecting the absence of allowed interband
absorption in the insulating regime.
At $\omega_0$, the onset of interband
transitions produces a rapid increase in
the conductivity, followed by a pronounced
resonance near the Van Hove frequency
$\hbar\omega_V\simeq2.73$~eV.
This enhancement originates from the
increased joint density of states near
saddle points of the electronic dispersion.
Additional spectral structures appear near
$\hbar\omega_Y=3.80$~eV and
$\hbar\omega_X=5.80$~eV, reflecting the
inequivalent interband transition energies
associated with the momentum-dependent
$d$-wave altermagnetic exchange.
The longitudinal conductivities of the
$C=\pm1$ phases nearly overlap, indicating
that reversing the uniform exchange field
leaves this response essentially unchanged.
In contrast, the $t_J=0$ reference exhibits
a shifted dominant resonance and lacks the
distinct $X/Y$ transition structure of
the altermagnetic system. The corresponding $\mathrm{Im}\,\sigma_{xx}(\omega)$ is displayed
in Fig.~\ref{fig:conductivity}(b).
Unlike its real counterpart, it remains finite
below the absorption threshold and develops
a pronounced negative dispersive feature
near $\omega_0$, consistent with the
Kramers--Kronig relations.
A rapid variation and sign change occur
near $\omega_V$, followed by additional
structures around $\omega_Y$ and $\omega_X$.
These features reflect the dispersive
counterparts of the interband absorption
structures observed in
$\mathrm{Re}\,\sigma_{xx}(\omega)$.
The nearly identical responses of the two
opposite Chern phases further demonstrate
the even character of the longitudinal
optical conductivity under exchange-field
reversal.

Figure~\ref{fig:conductivity}(c) illustrates
the real part of the transverse conductivity,
$\mathrm{Re}\,\sigma_{xy}(\omega)$, which exhibits
a qualitatively different dependence on the
topological phase.
In the insulating dc limit, the Hall response
is quantized according to
\begin{equation}
\mathrm{Re}\,\sigma_{xy}(0)
=
C\frac{e^2}{h},
\label{eq:dc_hall}
\end{equation}
yielding opposite Hall conductivities for
the $C=+1$ and $C=-1$ phases.
At finite photon energies, the response
develops a pronounced resonance near
$\omega_0$ and changes sign around
$\omega_V$, followed by additional
spectral features near $\omega_Y$ and
$\omega_X$.
Remarkably, the Hall conductivity reverses
its sign almost exactly between the two
Chern phases while retaining the same
spectral structure, satisfying
\begin{equation}
\sigma_{xy}(\omega;C=+1)
\simeq
-\sigma_{xy}(\omega;C=-1).
\label{eq:hall_reversal}
\end{equation}
In the absence of altermagnetic exchange
($t_J=0$), the Hall conductivity vanishes
throughout the investigated frequency
range, consistent with the trivial
topological character of the reference
system. Finally, Fig.~\ref{fig:conductivity}(d)
presents $\mathrm{Im}\,\sigma_{xy}(\omega)$,
which approaches zero in the static limit
and remains strongly suppressed below
the interband threshold.
A sharp onset appears near $\omega_0$,
followed by a pronounced enhancement
around $\omega_V$ and additional changes
near $\omega_Y$ and $\omega_X$.
As in the real Hall response, the imaginary
component exhibits opposite signs for
$C=+1$ and $C=-1$, whereas it vanishes
for $t_J=0$.

\subsection{Spectral signatures of the longitudinal and transverse splittings}
\begin{figure*}[t]
\centering
\includegraphics[width=\textwidth]{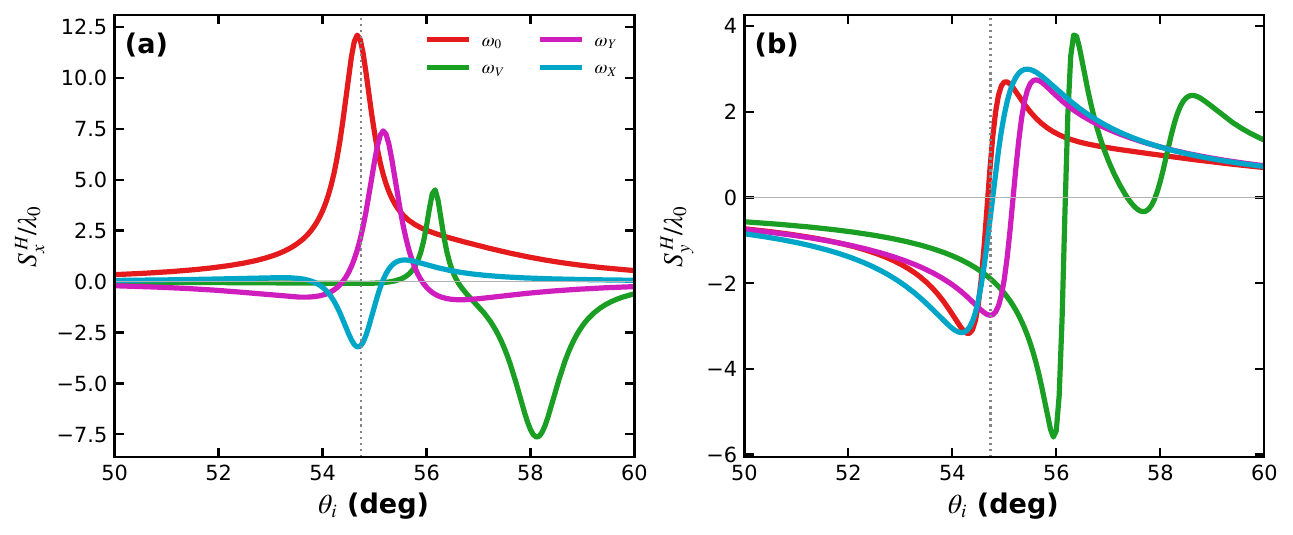}
\caption{\textcolor{black}{Angular variation of (a) $S_x^H/\lambda_0$ and (b) $S_y^H/\lambda_0$ at four characteristic photon energies. Red, green, magenta, and cyan curves denote $\omega_0$, $\omega_V$, $\omega_Y$, and $\omega_X$, respectively. These cuts use the same finite-beam data and parameters as Fig.~\ref{fig:shifts}.}}
\label{fig:angular}
\end{figure*}

Figure~\ref{fig:shifts}(a) presents the photon-energy
and incidence-angle dependence of the longitudinal
spin splitting $S_x^H/\lambda_0$ for the $C=+1$
altermagnetic phase. The response exhibits pronounced
positive and negative regions concentrated near the
pseudo-Brewster angle, with their angular positions
and magnitudes varying significantly across the
investigated photon-energy range. At a conducting
interface, the reflected field contains both
copolarized and cross-polarized components,
characterized by the Fresnel coefficients $r_{pp}$
and $r_{sp}$, respectively. The latter is directly
related to the optical Hall conductivity
$\sigma_{xy}(\omega)$ and therefore carries
information about the time-reversal-symmetry-breaking
electronic response. Near the pseudo-Brewster
minimum, the copolarized reflection amplitude is
strongly suppressed, increasing the relative
contribution of the Hall-induced cross-polarized
component and producing an enhanced longitudinal
spin splitting. The bare-substrate Brewster angle
is $\theta_B=\tan^{-1}(n_t/n_i)\simeq54.74^\circ$,
although the actual extrema are displaced from
this value by the complex and frequency-dependent
conductivity of the altermagnetic sheet.

The spectral structures in Fig.~\ref{fig:shifts}(a)
are closely associated with the characteristic
interband transition energies identified in the
optical conductivity. At the minimum excitation
threshold $\hbar\omega_0=1.00$~eV, the onset of
interband absorption modifies the reflection
amplitudes and produces a pronounced change in
the angular dependence of $S_x^H$. A particularly
strong spectral variation occurs near the
Van Hove energy $\hbar\omega_V\simeq2.73$~eV,
where the enhanced optical response associated
with saddle points of the electronic dispersion
is transferred to the reflected-beam shifts
through the complex Fresnel coefficients.
Additional structures appear near the
inequivalent interband transition energies
$\hbar\omega_Y=3.80$~eV and
$\hbar\omega_X=5.80$~eV. These transitions
reflect the momentum-dependent $d$-wave exchange,
which produces different electronic gaps at
the $X$ and $Y$ points. The resulting spectral
features thus connect the longitudinal spin
splitting to the underlying altermagnetic band
structure. Nevertheless, the magnitude and sign
of the shifts are determined jointly by the
optical conductivity, reflection amplitudes,
and their angular variations, rather than by
the joint density of states alone.

Figure~\ref{fig:shifts}(b) shows the corresponding
transverse spin splitting $S_y^H/\lambda_0$.
Unlike the longitudinal shift, the transverse
response exhibits a broad reversal of sign
across the pseudo-Brewster region, separating
positive and negative displacement domains.
This behavior arises from the geometric
spin-orbit interaction of light associated
with changes in the local plane of incidence
and is related to the Imbert--Fedorov effect.
The frequency-dependent optical conductivity
further modifies this geometric contribution,
leading to enhanced spectral structures near
$\omega_0$, $\omega_V$, $\omega_Y$, and $\omega_X$.
For the isotropic Hall tensor considered here,
reversal of $\sigma_{xy}$ changes the sign of
the cross-polarized reflection coefficients,
whereas the copolarized coefficients remain
unchanged. Consequently, the longitudinal
spin splitting $S_x^H$ is odd under Hall
reversal, while the transverse splitting
$S_y^H$ is even within the adopted polarization
and coordinate conventions. The two components
therefore provide complementary information
about the Hall-sensitive and geometric
contributions to the photonic spin Hall effect.

To resolve the angular characteristics of these
spectral features, Figs.~\ref{fig:angular}(a)
and \ref{fig:angular}(b) present the longitudinal
and transverse spin splittings at four selected
photon energies, respectively. Figure~\ref{fig:angular}(a)
reveals a strong dependence of $S_x^H$ on the
interband transition energy. The red curve,
corresponding to $\omega_0$, develops a pronounced
positive maximum near the pseudo-Brewster angle,
where the sensitivity to the Hall-induced
cross-polarized reflection amplitude is enhanced.
In contrast, the green curve at $\omega_V$
exhibits a sharp negative displacement followed
by a rapid angular variation, indicating that
the Van Hove-related optical response strongly
modifies the angular dependence of the
reflected field. The magenta and cyan curves,
corresponding to $\omega_Y$ and $\omega_X$,
respectively, exhibit additional extrema and
sign reversals associated with the inequivalent
interband transition channels. Their distinct
angular profiles illustrate how the
momentum-dependent altermagnetic exchange
influences the spin-dependent optical response
at different excitation energies.

The corresponding transverse displacements
are displayed in Fig.~\ref{fig:angular}(b).
All four curves undergo substantial changes
near the pseudo-Brewster region, but their
angular extrema and magnitudes differ
considerably. The response at $\omega_V$
shows a particularly sharp variation,
consistent with the enhanced dispersive
optical response near the Van Hove energy.
At the other characteristic energies,
the transverse splitting exhibits broader
variations and additional extrema arising
from the frequency-dependent balance
between the copolarized and cross-polarized
reflection amplitudes. These results
demonstrate that the longitudinal and
transverse spin splittings are governed
by different combinations of the Fresnel
coefficients and therefore probe distinct
aspects of the optical response.

\subsection{Low-frequency Chern-phase readout}
\begin{figure}[t]
\centering
\includegraphics[width=\columnwidth]{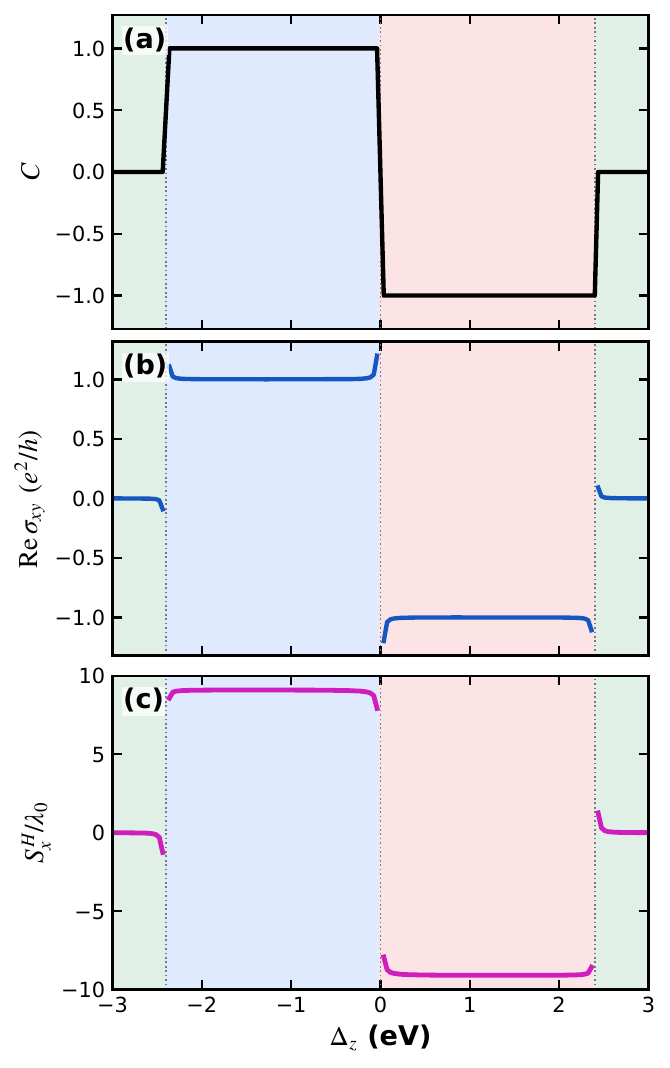}
\caption{
Low-frequency response as a function of the uniform
exchange field $\Delta_z$: (a) Chern number $C$,
(b) real Hall conductivity
$\mathrm{Re}\,\sigma_{xy}/(e^2/h)$, and
(c) normalized longitudinal spin splitting
$S_x^H/\lambda_0$.
Blue, pink, and green regions indicate the
$C=+1$, $C=-1$, and $C=0$ phases, respectively.
Vertical dotted lines mark the phase boundaries
$\Delta_z=0,\pm4t_J$.
Parameters are $t_J=0.6$~eV,
$\hbar\omega=0.05$~eV, and
$\theta_i=54.3^\circ$.
}
\label{fig:chern}
\end{figure}

\begin{figure}[t]
\centering
\includegraphics[width=\columnwidth]{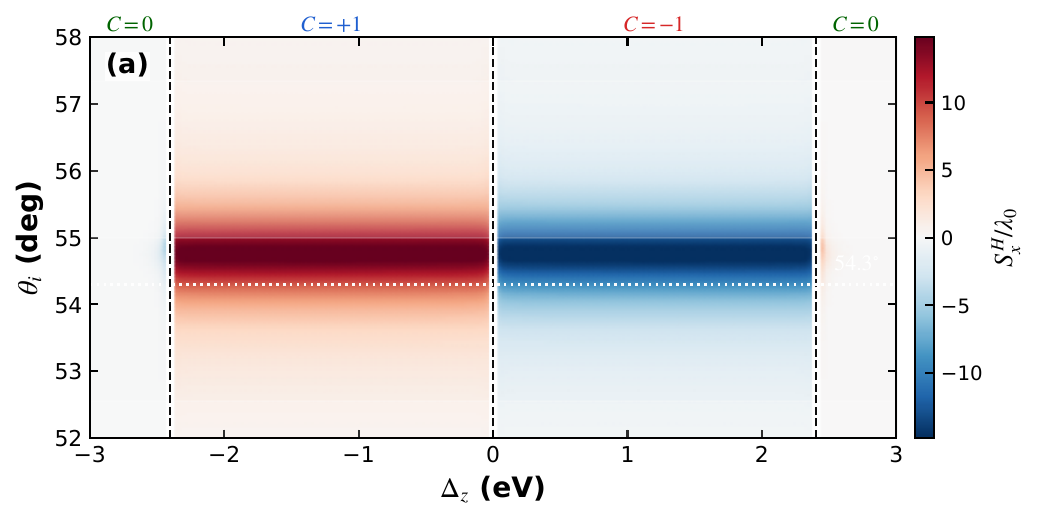}
\caption{
Exchange-field and incidence-angle dependence of
the finite-beam longitudinal spin splitting
$S_x^H/\lambda_0$ at $\hbar\omega=0.05$~eV
and $t_J=0.6$~eV.
Dashed vertical lines mark the phase boundaries,
and the horizontal dotted line indicates the
$\theta_i=54.3^\circ$ angular cut used in
Fig.~\ref{fig:chern}.
Other parameters are the same as in
Fig.~\ref{fig:chern}.
}
\label{fig:chernmap}
\end{figure}
To establish the relationship between the
photonic spin splitting and the electronic
topological phases, Fig.~\ref{fig:chern}(a)--(c)
presents the Chern number, the real part of
the optical Hall conductivity, and the
longitudinal spin splitting $S_x^H/\lambda_0$,
respectively, as functions of the uniform
exchange field $\Delta_z$ at
$\hbar\omega=0.05$~eV and $t_J=0.6$~eV.
As shown in Fig.~\ref{fig:chern}(a), the
topological phase transitions occur at
$\Delta_z=0$ and $\Delta_z=\pm4t_J
=\pm2.4$~eV, where the electronic gap
closes and reopens. The system exhibits
$C=+1$ for $-2.4<\Delta_z<0$~eV
and $C=-1$ for $0<\Delta_z<2.4$~eV,
whereas the outer regions are topologically
trivial ($C=0$). These phase boundaries
follow directly from the Dirac masses at
the high-symmetry points of the Brillouin
zone.

Figure~\ref{fig:chern}(b) shows the
corresponding low-frequency Hall conductivity,
$\mathrm{Re}\,\sigma_{xy}(\omega)$.
Within the well-gapped Chern-insulating
regions, the Hall response approaches
the quantized dc value $Ce^2/h$ and
reverses sign between the $C=+1$ and
$C=-1$ phases. Deviations from this
limiting behavior become pronounced
near the phase boundaries, where the
decreasing electronic gap enhances
finite-frequency effects. The correspondence
between the Chern number and the Hall
conductivity demonstrates how the
exchange-controlled band topology
is reflected in the optical response.

The resulting longitudinal spin splitting,
displayed in Fig.~\ref{fig:chern}(c),
exhibits a corresponding sign reversal.
In particular, $S_x^H$ is positive
in the $C=+1$ region and negative
in the $C=-1$ region, consistent
with its Hall-odd character in the
adopted reflection and polarization
conventions. For the representative
exchange fields $\Delta_z=\mp0.5$~eV,
the calculated splittings are
$S_x^H/\lambda_0\simeq\pm9.27$
at $\theta_i=54.3^\circ$.
By contrast, the transverse half-separation
$S_y^H$ remains essentially unchanged
under Hall reversal, reflecting its
Hall-even character in the isotropic
model. These results indicate that
the longitudinal splitting is particularly
sensitive to the reversal of the
exchange-controlled Hall response.
Its absolute sign, however, depends
on the definitions of the reflected
coordinates and circular polarization
components.

Figure~\ref{fig:chernmap} further illustrates
the exchange-field and incidence-angle
dependence of $S_x^H/\lambda_0$.
The map reveals positive and negative
displacement regions associated with
the opposite Chern phases, with the
largest responses concentrated near
the pseudo-Brewster region. The
approximately antisymmetric dependence
on $\Delta_z$ demonstrates that the
Hall-sensitive splitting persists
over a finite angular range rather
than at a single incidence angle.
The vertical phase boundaries at
$\Delta_z=0,\pm2.4$~eV correspond
to electronic gap closings and
should not be interpreted as optical
resonances. Near these boundaries,
the finite probe frequency and
spectral broadening limit the
applicability of the quantized
low-frequency Hall description.

To distinguish the influence of
$d$-wave altermagnetic exchange from
the generic optical response of a
Chern insulator, we also consider
a square-lattice reference model
described by
$\mathbf{d}_{\mathrm{ref}}
=(t\sin k_ya,-t\sin k_xa,
m+B[\cos k_xa+\cos k_ya])$.
For $t=1$~eV, $B=0.5$~eV, and
$m=0.5$~eV, this model exhibits
$C=+1$ with a minimum direct gap
of approximately $1.00$~eV,
matching the corresponding
altermagnetic phase. The reference
model produces a comparable
low-frequency Hall-sensitive
longitudinal splitting, indicating
that the existence of a Hall-odd
photonic spin displacement is not
unique to altermagnets. Instead,
the distinguishing features of
the present model arise from the
momentum-dependent $d$-wave
exchange, which generates
inequivalent $X$- and $Y$-point
interband transition energies
and modifies the frequency-dependent
conductivity and spin-splitting spectra.
For the separate $t_J=0$ reference
configuration considered here,
the vanishing optical Hall
conductivity also yields
$S_x^H\approx0$, while the
Hall-even transverse response
remains finite. 

\section{Conclusions}
\label{sec:conclusions}
We investigated the photonic spin Hall effect
in a Rashba-coupled $d$-wave altermagnetic Chern
model. The interplay between altermagnetic and
uniform exchange fields produced distinct
topological phases characterized by Chern
numbers $C=0,\pm1$. The calculated optical
conductivities revealed characteristic
interband transitions, a Van Hove spectral
enhancement, and inequivalent $X$- and
$Y$-point transition energies, which were
reflected in the frequency-dependent spin dependent splittings. We found that the longitudinal spin splitting
$S_x^H$ reversed sign between the $C=+1$
and $C=-1$ phases, following the reversal
of the optical Hall conductivity, whereas
the transverse splitting $S_y^H$ remained
Hall even within the adopted model.
Both responses exhibited pronounced
enhancement near the pseudo-Brewster region.
The sensitivity of the longitudinal splitting
to Hall-conductivity reversal provides a
direct connection between the electronic
topological phases and the polarization-dependent
reflection response. Furthermore, the distinct
spectral features associated with the $d$-wave
exchange demonstrated how the underlying
electronic structure influences the spatial
separation of the reflected circular components. These findings highlight the potential of
polarization-resolved photonic spin Hall shifts
for probing the interplay between electronic
topology, altermagnetic exchange, and optical
Hall responses. The combined spectral and
angular characteristics offer complementary
signatures of the exchange-controlled
electronic phases and provide a theoretical
basis for future optical investigations
of topological altermagnetic systems.


\bibliographystyle{apsrev4-2}
\bibliography{references}

@article{Smejkal2022PRX,
  author = {L. \v{S}mejkal and J. Sinova and T. Jungwirth},
  title = {{Beyond conventional ferromagnetism and antiferromagnetism: A phase with nonrelativistic spin and crystal rotation symmetry}},
  journal = {Phys. Rev. X},
  volume = {12},
  pages = {031042},
  year = {2022},
}

@article{Smejkal2022Landscape,
  author = {L. \v{S}mejkal and J. Sinova and T. Jungwirth},
  title = {{Emerging research landscape of altermagnetism}},
  journal = {Phys. Rev. X},
  volume = {12},
  pages = {040501},
  year = {2022},
}

@article{Mazin2022,
  author = {{I. Mazin (The PRX Editors)}},
  title = {{Editorial: Altermagnetism---A new punch line of fundamental magnetism}},
  journal = {Phys. Rev. X},
  volume = {12},
  pages = {040002},
  year = {2022},
}

@article{Hayami2019,
  author = {S. Hayami and Y. Yanagi and H. Kusunose},
  title = {{Momentum-dependent spin splitting by collinear antiferromagnetic ordering}},
  journal = {J. Phys. Soc. Jpn.},
  volume = {88},
  pages = {123702},
  year = {2019},
}

@article{Bai2024,
  author = {L. Bai and W. Feng and S. Liu and L. \v{S}mejkal and Y. Mokrousov and Y. Yao},
  title = {{Altermagnetism: Exploring new frontiers in magnetism and spintronics}},
  journal = {Adv. Funct. Mater.},
  volume = {34},
  pages = {2409327},
  year = {2024},
}

@article{Song2025,
  author = {C. Song and H. Bai and Z. Zhou and L. Han and H. Reichlova and J. H. Dil and J. Liu and X. Chen and F. Pan},
  title = {{Altermagnets as a new class of functional materials}},
  journal = {Nat. Rev. Mater.},
  volume = {10},
  pages = {473},
  year = {2025},
}

@article{Krempasky2024,
  author = {J. Krempask\'y and L. \v{S}mejkal and S. W. D'Souza and M. Hajlaoui and G. Springholz and K. Uhl\'{\i}\v{r}ov\'a and F. Alarab and P. C. Constantinou and V. Strocov and D. Usanov and others},
  title = {{Altermagnetic lifting of Kramers spin degeneracy}},
  journal = {Nature},
  volume = {626},
  pages = {517},
  year = {2024},
}

@article{Lee2024,
  author = {S. Lee and S. Lee and S. Jung and J. Jung and D. Kim and Y. Lee and B. Seok and J. Kim and B. G. Park and L. \v{S}mejkal and C.-J. Kang and C. Kim},
  title = {{Broken Kramers degeneracy in altermagnetic MnTe}},
  journal = {Phys. Rev. Lett.},
  volume = {132},
  pages = {036702},
  year = {2024},
}

@article{Osumi2024,
  author = {T. Osumi and S. Souma and T. Aoyama and K. Yamauchi and A. Honma and K. Nakayama and T. Takahashi and K. Ohgushi and T. Sato},
  title = {{Observation of a giant band splitting in altermagnetic MnTe}},
  journal = {Phys. Rev. B},
  volume = {109},
  pages = {115102},
  year = {2024},
}

@article{Reimers2024,
  author = {S. Reimers and L. Odenbreit and L. \v{S}mejkal and V. N. Strocov and P. Constantinou and A. B. Hellenes and R. Jaeschke Ubiergo and W. H. Campos and V. K. Bharadwaj and A. Chakraborty and others},
  title = {{Direct observation of altermagnetic band splitting in CrSb thin films}},
  journal = {Nat. Commun.},
  volume = {15},
  pages = {2116},
  year = {2024},
}

@article{Ding2024,
  author = {J. Ding and Z. Jiang and X. Chen and Z. Tao and Z. Liu and T. Li and J. Liu and J. Sun and J. Cheng and J. Liu and others},
  title = {{Large band splitting in $g$-wave altermagnet CrSb}},
  journal = {Phys. Rev. Lett.},
  volume = {133},
  pages = {206401},
  year = {2024},
}

@article{Jiang2025,
  author = {B. Jiang and M. Hu and J. Bai and Z. Song and C. Mu and G. Qu and W. Li and W. Zhu and H. Pi and Z. Wei and others},
  title = {{A metallic room-temperature $d$-wave altermagnet}},
  journal = {Nat. Phys.},
  volume = {21},
  pages = {754},
  year = {2025},
}

@article{GonzalezBetancourt2023,
  author = {R. D. Gonzalez Betancourt and J. Zub\'a\v{c} and R. Gonzalez-Hernandez and K. Geishendorf and Z. \v{S}ob\'a\v{n} and G. Springholz and K. Olejn\'{\i}k and L. \v{S}mejkal and J. Sinova and T. Jungwirth and S. T. B. Goennenwein and A. Thomas and H. Reichlov\'a and J. \v{Z}elezn\'y and D. Kriegner},
  title = {{Spontaneous anomalous Hall effect arising from an unconventional compensated magnetic phase in a semiconductor}},
  journal = {Phys. Rev. Lett.},
  volume = {130},
  pages = {036702},
  year = {2023},
}

@article{Reichlova2024,
  author = {H. Reichlova and R. Lopes Seeger and R. Gonz\'alez-Hern\'andez and I. Kounta and R. Schlitz and D. Kriegner and P. Ritzinger and M. Lammel and M. Leivisk\"a and A. Birk Hellenes and others},
  title = {{Observation of a spontaneous anomalous Hall response in the Mn$_5$Si$_3$ $d$-wave altermagnet candidate}},
  journal = {Nat. Commun.},
  volume = {15},
  pages = {4961},
  year = {2024},
}

@article{Ma2024,
  author = {H.-Y. Ma and J.-F. Jia},
  title = {{Altermagnetic topological insulator and the selection rules}},
  journal = {Phys. Rev. B},
  volume = {110},
  pages = {064426},
  year = {2024},
}

@article{LiLiu2024,
  author = {Y.-X. Li and Y. Liu and C.-C. Liu},
  title = {{Creation and manipulation of higher-order topological states by altermagnets}},
  journal = {Phys. Rev. B},
  volume = {109},
  pages = {L201109},
  year = {2024},
}

@article{LiZhang2025,
  author = {Y.-Y. Li and S.-B. Zhang},
  title = {{Floating edge bands in the Bernevig-Hughes-Zhang model with altermagnetism}},
  journal = {Phys. Rev. B},
  volume = {111},
  pages = {045106},
  year = {2025},
}

@article{Antonenko2025,
  author = {D. S. Antonenko and R. M. Fernandes and J. W. F. Venderbos},
  title = {{Mirror Chern bands and Weyl nodal loops in altermagnets}},
  journal = {Phys. Rev. Lett.},
  volume = {134},
  pages = {096703},
  year = {2025},
}

@article{Parshukov2025,
  author = {K. Parshukov and R. Wiedmann and A. P. Schnyder},
  title = {{Topological crossings in two-dimensional altermagnets: Symmetry classification and topological responses}},
  journal = {Phys. Rev. B},
  volume = {111},
  pages = {224406},
  year = {2025},
}

@article{GonzalezHernandez2025,
  author = {R. Gonz\'alez-Hern\'andez and H. Serrano and B. Uribe},
  title = {{Spin Chern number in altermagnets}},
  journal = {Phys. Rev. B},
  volume = {111},
  pages = {085127},
  year = {2025},
}

@article{Rao2024,
  author = {P. Rao and A. Mook and J. Knolle},
  title = {{Tunable band topology and optical conductivity in altermagnets}},
  journal = {Phys. Rev. B},
  volume = {110},
  pages = {024425},
  year = {2024},
}

@article{Cheraghchi2026,
  author = {H. Cheraghchi},
  title = {{Floquet-engineered Chern insulator in two-dimensional $d_{x^2-y^2}$-wave altermagnets}},
  journal = {Phys. Rev. B},
  volume = {113},
  pages = {155439},
  year = {2026},
  doi = {10.1103/mslj-zz6t},
}

@misc{Qin2026,
  author = {Fang Qin and Rui Chen and Xiao-Bin Qiang},
  title = {{Optical conductivity signature of Van Hove singularity in altermagnetic topological systems}},
  year = {2026},
  eprint = {2608.20986},
  archivePrefix = {arXiv},
  url = {https://arxiv.org/abs/2608.20986}
}

@article{Tse2010,
  author = {W.-K. Tse and A. H. MacDonald},
  title = {{Giant magneto-optical Kerr effect and universal Faraday effect in thin-film topological insulators}},
  journal = {Phys. Rev. Lett.},
  volume = {105},
  pages = {057401},
  year = {2010},
}

@article{Tse2011,
  author = {W.-K. Tse and A. H. MacDonald},
  title = {{Magneto-optical Faraday and Kerr effects in topological insulator films and in other layered quantized Hall systems}},
  journal = {Phys. Rev. B},
  volume = {84},
  pages = {205327},
  year = {2011},
}

@article{Chen2025,
  author = {W. Chen and X. Zhou and W.-K. Lou and K. Chang},
  title = {{Magneto-optical conductivity and circular dichroism in $d$-wave altermagnets}},
  journal = {Phys. Rev. B},
  volume = {111},
  pages = {064428},
  year = {2025},
}

@article{Iguchi2025,
  author = {S. Iguchi and H. Kobayashi and Y. Ikemoto and T. Furukawa and H. Itoh and S. Iwai and T. Moriwaki and T. Sasaki},
  title = {{Magneto-optical spectra of an organic antiferromagnet as a candidate for an altermagnet}},
  journal = {Phys. Rev. Res.},
  volume = {7},
  pages = {033026},
  year = {2025},
}

@misc{LiLiu2025,
  author = {Yongpan Li and Yichen Liu and Cheng-Cheng Liu},
  title = {{Unconventional magneto-optical effects in altermagnets}},
  year = {2025},
  eprint = {2512.03435},
  archivePrefix = {arXiv},
  url = {https://arxiv.org/abs/2512.03435}
}

@misc{Tagani2026,
  author = {M. B. Tagani and C. Autieri and W. Brzezicki},
  title = {{Band topology and symmetry-driven magneto-optical response in two-dimensional $d$-wave altermagnets with staggered spin-orbit coupling}},
  year = {2026},
  eprint = {2608.02155},
  archivePrefix = {arXiv},
  url = {https://arxiv.org/abs/2608.02155}
}

@article{Onoda2004,
  author = {M. Onoda and S. Murakami and N. Nagaosa},
  title = {{Hall effect of light}},
  journal = {Phys. Rev. Lett.},
  volume = {93},
  pages = {083901},
  year = {2004},
}

@article{Bliokh2006,
  author = {K. Y. Bliokh and Y. P. Bliokh},
  title = {{Conservation of angular momentum, transverse shift, and spin Hall effect in reflection and refraction of an electromagnetic wave packet}},
  journal = {Phys. Rev. Lett.},
  volume = {96},
  pages = {073903},
  year = {2006},
}

@article{Hosten2008,
  author = {O. Hosten and P. Kwiat},
  title = {{Observation of the spin Hall effect of light via weak measurements}},
  journal = {Science},
  volume = {319},
  pages = {787},
  year = {2008},
}

@article{Bliokh2015,
  author = {K. Y. Bliokh and F. J. Rodr\'{\i}guez-Fortu\~no and F. Nori and A. V. Zayats},
  title = {{Spin--orbit interactions of light}},
  journal = {Nat. Photonics},
  volume = {9},
  pages = {796},
  year = {2015},
}

@article{Ling2017,
  author = {X. Ling and X. Zhou and K. Huang and Y. Liu and C.-W. Qiu and H. Luo and S. Wen},
  title = {{Recent advances in the spin Hall effect of light}},
  journal = {Rep. Prog. Phys.},
  volume = {80},
  pages = {066401},
  year = {2017},
}

@article{Luo2011,
  author = {H. Luo and X. Zhou and W. Shu and S. Wen and D. Fan},
  title = {{Enhanced and switchable spin Hall effect of light near the Brewster angle on reflection}},
  journal = {Phys. Rev. A},
  volume = {84},
  pages = {043806},
  year = {2011},
}

@article{Kong2012,
  author = {L.-J. Kong and X.-L. Wang and S.-M. Li and Y. Li and J. Chen and B. Gu and H.-T. Wang},
  title = {{Spin Hall effect of reflected light from an air-glass interface around the Brewster's angle}},
  journal = {Appl. Phys. Lett.},
  volume = {100},
  pages = {071109},
  year = {2012},
}

@article{Zhou2012,
  author = {X. Zhou and X. Ling and H. Luo and S. Wen},
  title = {{Identifying graphene layers via spin Hall effect of light}},
  journal = {Appl. Phys. Lett.},
  volume = {101},
  pages = {251602},
  year = {2012},
}

@article{Cai2017,
  author = {L. Cai and M. Liu and S. Chen and Y. Liu and W. Shu and H. Luo and S. Wen},
  title = {{Quantized photonic spin Hall effect in graphene}},
  journal = {Phys. Rev. A},
  volume = {95},
  pages = {013809},
  year = {2017},
}

@article{KortKamp2017,
  author = {W. J. M. Kort-Kamp},
  title = {{Topological phase transitions in the photonic spin Hall effect}},
  journal = {Phys. Rev. Lett.},
  volume = {119},
  pages = {147401},
  year = {2017},
}

@article{Shah2022,
  author = {M. Shah},
  title = {{Probing topological quantum phase transitions via photonic spin Hall effects in spin-orbit coupled 2D quantum materials}},
  journal = {J. Phys. D: Appl. Phys.},
  volume = {55},
  pages = {105105},
  year = {2022},
}

@article{Shah2024,
  author = {M. Shah and M. S. Anwar and R. Asgari and G. Xianlong},
  title = {{Photonic spin Hall effect in Haldane model materials}},
  journal = {Phys. Rev. B},
  volume = {109},
  pages = {235418},
  year = {2024},
}

@misc{Samanta2026,
  author = {Deblina Samanta and Darshan G. Joshi},
  title = {{Photonic spin-Hall effect as a probe for time-reversal-symmetry broken band topological phases}},
  year = {2026},
  eprint = {2608.25002},
  archivePrefix = {arXiv},
  url = {https://arxiv.org/abs/2608.25002}
}

@article{BliokhAiello2013,
  author = {K. Y. Bliokh and A. Aiello},
  title = {{Goos--H\"anchen and Imbert--Fedorov beam shifts: An overview}},
  journal = {J. Opt.},
  volume = {15},
  pages = {014001},
  year = {2013},
}

@article{RevModPhys.82.1539,
  title = {Anomalous Hall effect},
  author = {Nagaosa, Naoto and Sinova, Jairo and Onoda, Shigeki and MacDonald, A. H. and Ong, N. P.},
  journal = {Rev. Mod. Phys.},
  volume = {82},
  issue = {2},
  pages = {1539--1592},
  numpages = {0},
  year = {2010},
  month = {May},
  publisher = {American Physical Society},
  doi = {10.1103/RevModPhys.82.1539},
  url = {https://link.aps.org/doi/10.1103/RevModPhys.82.1539}
}

@article{KortKamp2016,
  author = {W. J. M. Kort-Kamp and N. A. Sinitsyn and D. A. R. Dalvit},
  title = {{Quantized beam shifts in graphene}},
  journal = {Phys. Rev. B},
  volume = {93},
  pages = {081410},
  year = {2016},
}

\end{document}